\documentclass[twocolumn,prb,citeautoscript,superscriptaddress,8pt]{revtex4-2}

\usepackage{graphicx}
\usepackage{gensymb}
\usepackage{color}
\usepackage[dvipsnames]{xcolor}
\usepackage{float}
\usepackage{upgreek}
\usepackage{bm}
\usepackage{amsmath,amssymb}

\begin{document}

\title{A compact, portable device for microscopic magnetic imaging based on diamond quantum sensors}

\author{Alex Shaji} % S3961863@student.rmit.edu.au
\affiliation{School of Science, RMIT University, Melbourne, VIC 3001, Australia}

\author{Kevin J. Rietwyk} % kevin.rietwyk@rmit.edu.au
\affiliation{School of Science, RMIT University, Melbourne, VIC 3001, Australia}

\author{Islay O. Robertson} % S3930288@student.rmit.edu.au
\affiliation{School of Science, RMIT University, Melbourne, VIC 3001, Australia}

\author{Philipp Reineck} % philipp.reineck@rmit.edu.au
\affiliation{School of Science, RMIT University, Melbourne, VIC 3001, Australia}
\affiliation{ARC Centre of Excellence for Nanoscale BioPhotonics, School of Science, RMIT University, Melbourne, VIC 3001, Australia}

\author{David A. Broadway} % david.broadway@rmit.edu.au
\email{david.broadway@rmit.edu.au}
\affiliation{School of Science, RMIT University, Melbourne, VIC 3001, Australia}

\author{Jean-Philippe Tetienne}
\email{jean-philippe.tetienne@rmit.edu.au}
\affiliation{School of Science, RMIT University, Melbourne, VIC 3001, Australia}

\begin{abstract} 

Magnetic imaging based on ensembles of diamond nitrogen-vacancy quantum sensors has emerged as a useful technique for the spatial characterisation of magnetic materials and current distributions. However, demonstrations have so far been restricted to laboratory-based experiments using relatively bulky apparatus and requiring manual handling of the diamond sensing element, hampering broader adoption of the technique. Here we present a simple, compact device that can be deployed outside a laboratory environment and enables robust, simplified operation. It relies on a specially designed sensor head that directly integrates the diamond sensor while incorporating a microwave antenna and all necessary optical components. This integrated sensor head is complemented by a small control unit and a laptop computer that displays the resulting magnetic image. We test the device by imaging a magnetic sample, demonstrating a spatial resolution of 4 $\mu$m over a field of view exceeding 1 mm, and a best sensitivity of 45 $\mu$T/$\sqrt{\rm Hz}$ per ($5\,\mu$m)$^2$ pixel. Our portable magnetic imaging instrument may find use in situations where taking the sample to be measured to a specialist lab is impractical or undesirable. 

\end{abstract}

\maketitle 

Widefield quantum diamond microscopy (QDM), which utilises a layer of nitrogen-vacancy (NV) centres in a diamond chip to image the magnetic fields emanating from a proximal sample, has emerged as a useful technique for quantitative, spatially resolved characterisation of magnetic materials and current distributions \cite{Steinert2010,Levine2019,Scholten2021}. Key features of QDM include a large field of view (up to several millimetres), a sub-micron spatial resolution, a high magnetic sensitivity, and operation under ambient conditions. These features have enabled the detection and precise imaging of magnetic signals in a variety of contexts, including magnetic inclusions in rocks~\cite{Glenn2017,Fu2020,Steele2023}, solid-state ferromagnets~\cite{Toraille2018,Broadway2020,Meirzada2021,Huang2023,Chen2022,Lamichhane2023}, biological organisms and tissues~\cite{LeSage2013,Fescenko2019,Mccoey2020,Chen2023}, and electric currents in integrated circuits~\cite{Turner2020,Kehayias2023,Garsi2024,Wen2024} and photovoltaic devices~\cite{Scholten2022}, to name just a few applications. 

One attractive aspect of QDM is its relative simplicity and potential for compactness and low cost. Indeed, there are no moving parts and the primary components required are simply a diamond containing an NV layer, a light source to optically excite the NV layer, a camera to image the red photoluminescence (PL), and a microwave (MW) source to drive the NVs' spin resonances, allowing optically detected magnetic resonance (ODMR) spectra to be collected and spatially correlated. This makes QDM a promising alternative to established magnetic imaging techniques such as magneto-optical Kerr microscopy, magnetic force microscopy, and Lorentz electron microscopy~\cite{Freeman2001}, which are typically costly to own and operate and require a permanent, specialised laboratory space. In contrast, it should be possible to realise a QDM system sufficiently compact and light-weight to be portable, in the sense that it could be routinely moved by the user according to their needs and space restrictions, and even deployed outside a laboratory environment. This would make applications such as rapid on-site screening of mineral samples or quality control of integrated circuits possible. However, such a portable QDM system has not been demonstrated yet. Furthermore, existing QDM systems are typically cumbersome to operate due to the need to manually place and align the diamond sensor on the sample to be imaged~\cite{Scholten2021}, hindering routine use by non-specialists.  

\begin{figure}[t!]
    \centering
    \includegraphics[width=8.5cm]{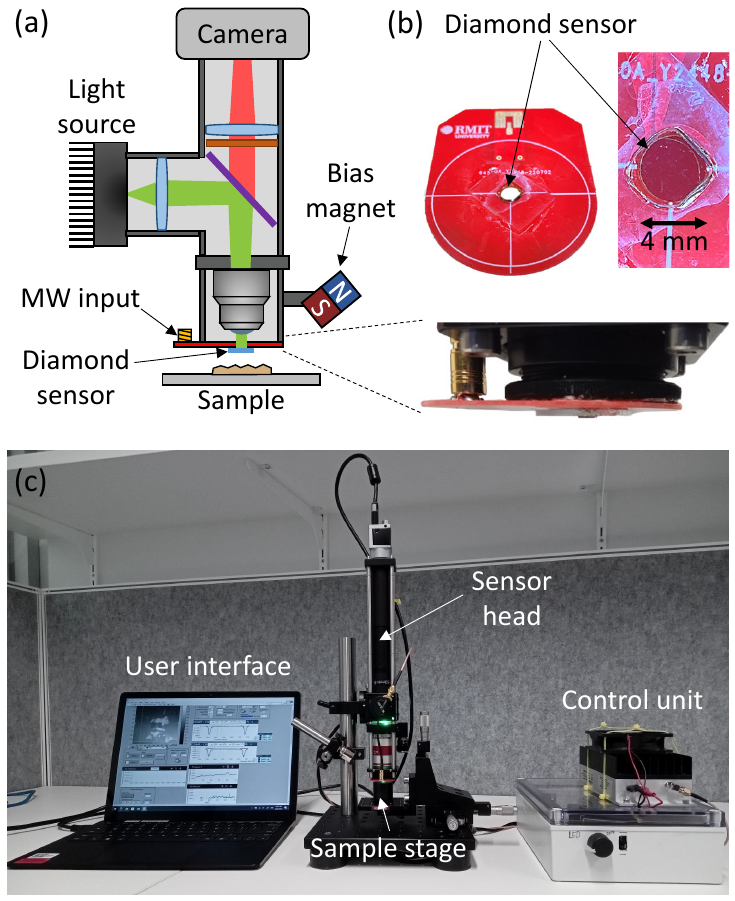}
    \caption{(a) Schematic of the sensor head of the compact quantum diamond microscope. The magnetically sensitive diamond is placed at the bottom of the head, directly facing the sample to be imaged which is mounted on a separate stage. (b) Photographs of the PCB holding the diamond sensor and integrating a MW loop antenna. (c) Photograph of the complete system, comprising a control unit and a laptop computer.}
    \label{fig1}
\end{figure}

In this letter, we report on a QDM device specially designed to be compact, robust, portable, and easy to operate without specialist expertise. Our device assembly relies on a mechanically fixed diamond sensor, thus removing the need for manual handling. All the components necessary for operation are incorporated either in the sensor head or in a small supporting control unit, making the overall system portable. We characterise the performance of our device (field of view, spatial resolution, magnetic sensitivity) and compare two versions differing by the choice of the light source. Future improvements in performance and portability are discussed.  

A schematic of the sensor head is presented in Fig.~\ref{fig1}(a). It comprises the light source, the camera, an objective lens and other basic optical components (collimation lens, tube lens, dichroic mirror, emission filter), and a permanent magnet used to bias the spin resonances of the NVs. The diamond sensor is attached to the head via a printed circuit board (PCB) equipped with a MW loop antenna to drive the spin resonances. The photographs in Fig.~\ref{fig1}(b) show an example realisation of the PCB-diamond assembly, where here a 4\,mm\,$\times$\,4\,mm$\times$\,400\,$\mu$m diamond glued on a supporting glass slide covers a 4-mm-diameter hole made in the PCB to allow light to go through, with the MW loop concentric to the hole. 

A photograph of the complete setup is shown in Fig.~\ref{fig1}(c). The sensor head occupies a volume of $\sim(10$\,cm$)^3$ and is mounted on a 20\,cm\,$\times$\,20\,cm breadboard. The sample to be imaged sits on a 3-axis translation stage placed directly under the sensor head, allowing the sample to be brought in contact with the diamond. Apart from the custom PCB and diamond, the sensor head is entirely built from commercially available off-the-shelf components (primarily sourced from Thorlabs) using a cage system. The complete list of components used is given in the Supplementary Material. The control unit [volume $\sim(20$\,cm$)^3$] encloses a MW signal generator (Windfreak SynthNV Pro) and amplifier (Mini-Circuits ZHL-42), a controller for the light source, and a 30\,W power supply. The weight of the sensor head (including sample stage and breadboard) and control unit is 4.9 and 2.3\,kg, respectively, making the system highly portable. A laptop computer connected to the control unit (to control the MW generator) and the sensor head (to control the camera) runs a custom LabVIEW program serving as the user interface, performing data acquisition and basic analysis. Advanced data processing to generate a magnetic field image is carried out via a separate Python script.

\begin{figure}[t!]
    \centering
    \includegraphics[width=8cm]{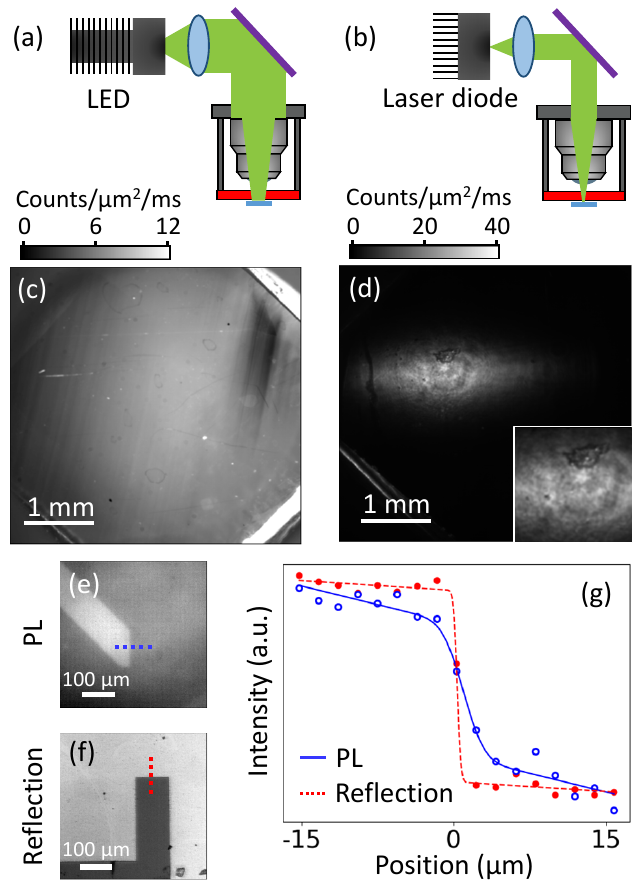}
    \caption{(a,b) Schematics of the sensor head equipped with either (a) a light emitting diode (LED) or (b) a laser diode as the light source. (c,d) PL images of the NV layer using (c) the LED and (d) the laser diode. The inset in (d) is a magnified view of the centre of the laser spot (1\,mm\,$\times$\,1\,mm). (e) PL image of an NV layer in contact with a metallic marker, used to estimate the optical resolution under QDM imaging conditions. (f) Reflection image (using a blue LED at 455 nm wavelength) of a metallic marker on a glass slide, used to estimate the raw optical resolution of the system (i.e., without diamond). (g) Line profiles extracted from (e) and (f) taken along the dotted lines. The two curves are independently re-scaled along the $y$-axis to facilitate comparison.}
    \label{fig2} 
\end{figure}

A critical component of a QDM setup is the green light source, which needs to provide sufficient power density over the desired field of view in order to efficiently excite the NV layer. In the literature this is typically achieved via a high-power ($\sim1$\,W) diode-pumped solid-state laser, but such lasers tend to be relatively heavy, bulky, and costly, all of which are undesirable features for our portable QDM system. We therefore explored two alternative light sources which can both be directly integrated into the sensor head: a light-emitted diode (LED, Thorlabs M530L4), and a laser diode (LD, Thorlabs L520P50 with mount LDM9T). The collimated light from the chosen light source was passed through a $f=50$\,mm objective lens (Thorlabs TL4X-SAP, 4x, 0.2 NA), which was also used to collect the PL. The LED outputs 460 mW of power, but high emission divergence and finite emitter size limit the maximum power density achievable at the NV layer due to coupling losses (only 160 mW make it to the diamond surface) and a large minimum spot size ($\sim2.5$\,mm), as illustrated in Fig.~\ref{fig2}(a). On the other hand, the LD outputs only 50 mW but can be focused on the NV layer with arbitrary spot size and minimal losses (40 mW at the diamond) [Fig.~\ref{fig2}(b)].

To illustrate the pros and cons of each light source, we recorded PL images [Fig.~\ref{fig2}(c,d)] of a diamond mounted to the sensor head as shown in Fig.~\ref{fig1}(b). The diamonds used in this work are high-pressure high-temperature (HPHT) grown single-crystal substrates in which a 500-nm-thick NV layer was formed at the bottom surface via ion implantation and annealing~\cite{Huang2013,Healey2020}. Given the pixel size of the camera (Basler acA2040-90um USB3 Mono) and $f=150$\,mm tube lens used, the images are 3.7\,mm\,$\times$\,3.7\,mm in size. With the collimated LED [Fig.~\ref{fig2}(c)], the illumination spot size is $\approx2.5$\,mm as expected (full width at half maximum, FWHM), which allows most of the diamond to be utilised for imaging though this comes at the cost of a relatively low power density ($\sim10$\,W/cm$^2$). With the LD [Fig.~\ref{fig2}(d)], the desired spot size can be obtained by adjusting the collimation lens, here the spot is elliptical with FWHM of 1.9 and 0.5\,mm along its principal axes. Note that the spot could easily be made circular by adding cylindrical lenses. At the centre of the LD spot the power density is more than 3 times larger than in the LED case, as indicated by the PL counts, despite the 4 times lower power. However, a downside of the LD is the poorer clarity of the PL image due to interference effects, contrasting with the much cleaner LED image.    

To characterise the spatial resolution of the optical setup, we placed a similar diamond in contact (NV face down) with a metallic marker, providing us with sharp boundaries in the PL image of the NV layer [Fig.~\ref{fig2}(e)]. This image was compared with a reflection image of a metallic marker on a glass slide, without the diamond present [Fig.~\ref{fig2}(f)].  
Line profiles across boundaries [see examples in Fig.~\ref{fig2}(g)] were fitted with an error function, i.e. the convolution of a step function with a Gaussian distribution. The FWHM of the Gaussian is thus a measure of the spatial resolution. In the reflection image, the FWHM is $1.6\pm0.9\,\mu$m, consistent with the diffraction limit. In the NV PL image, however, it is found to be $3.9\pm1.2\,\mu$m, larger than the diffraction limit ($\approx1.8\,\mu$m with the objective used at 700\,nm wavelength), which is attributed due to the finite NV layer thickness as well as additional diamond-related optical aberrations~\cite{Nishimura2024}. This FWHM of about $4\,\mu$m thus corresponds to the optical resolution of the system in QDM imaging conditions.

\begin{figure*}[t!]
    \centering
    \includegraphics[width=15cm]{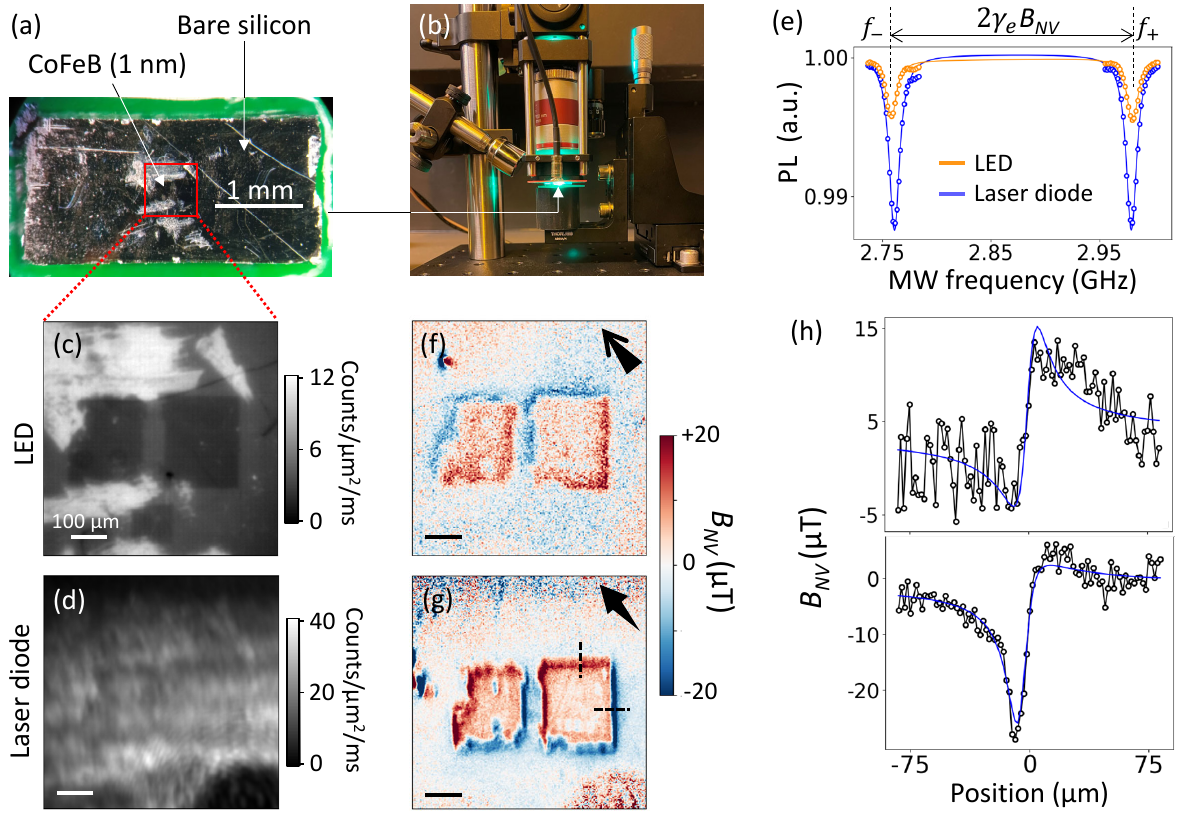}
    \caption{(a) Photograph of the magnetic test sample. (b) Photograph of the system during data acquisition. (c,d) PL images of the NV layer with the sample in contact, using (c) the LED or (d) the LD as the excitation source. (e) ODMR spectrum integrated over the entire region imaged in (c,d). The solid lines are double Lorentzian fits. (f,g) Maps of the magnetic field projection $B_{\rm NV}$ corresponding to the PL image in (c,d). N.B: the NV projection axis (set by the external magnet) is different in (f) and (g), as indicated by the black arrow in each image. (h) Line profiles along the vertical (top graph) and horizontal (bottom graph) dashed lines in (g). The solid lines are a fit to a model (see text).}
    \label{fig3}
\end{figure*}

We now test the ability of our portable QDM system to perform magnetic imaging. As a test sample, we used a perpendicularly magnetised CoFeB film (1\,nm thick) patterned on a Si substrate, pictured in Fig.~\ref{fig3}(a). The sample was placed on a rubber foot (Thorlabs AV4) sitting on the translation stage, and brought up until full contact with the diamond was achieved [Fig.~\ref{fig3}(b)]. The PL image of the NV layer obtained with the LED is shown in Fig.~\ref{fig3}(c), in which sample features are clearly visible. We note that the optical focus was fine-tuned after contact with the sample was made, by adjusting the vertical translation stage on which the objective is mounted (Thorlabs SM1ZA), which moves the objective relative to the diamond. When the LD is used instead of the LED, the PL image becomes dominated by interference effects and other illumination inhomogeneities [Fig.~\ref{fig3}(d)], making it much harder to locate the region of interest in the sample.

Continuous-wave ODMR spectra were acquired by collecting PL images while sweeping the MW frequency. A small bias magnetic field $B_0\approx8$\,mT was applied along one NV orientation family of the (100)-oriented diamond. This allows us to isolate the two spin resonances (at frequencies $f_\pm$) of this family in the ODMR spectrum, as shown in Fig.~\ref{fig3}(e), here corresponding to the PL from the entire region imaged in Fig.~\ref{fig3}(c,d). The typical exposure time per MW frequency was 80\,ms (chosen to approach camera saturation) and a MW-off reference frame was taken between each frequency step for normalisation, resulting in a sweep time of less than a minute. Sweeps were repeated to accumulate signal until the desired signal-to-noise ratio was reached. As seen in Fig.~\ref{fig3}(e), the ODMR contrast is roughly 3 times higher with the LD ($\approx1.2\%$) compared to the LED ($\approx0.4\%$), which is due to the larger excitation power density in the LD case.

A magnetic field map is obtained by fitting the ODMR spectrum to extract the frequencies $f_\pm$ at each pixel and converting to a magnetic field $B_{\rm NV}=(f_+-f_-)/2\gamma_e$. Here $\gamma_e=28$\,MHz/mT is the electron gyromagnetic ratio and $B_{\rm NV}$ is the magnetic field projection along the selected NV axis~\cite{Scholten2021}. The resulting $B_{\rm NV}$ maps obtained after fourteen hours of integration are shown in Fig.~\ref{fig3}(f,g) for the LED and LD cases, respectively. The characteristic stray field from the perpendicularly magnetised structures is clearly resolved in both cases, but the noise level is significantly lower with the LD. The magnetic sensitivity deduced from the pixel-to-pixel noise in these images is 224 and 45\,$\mu$T/$\sqrt{\rm Hz}$ with the LED and LD, respectively, per ($5\,\mu$m)$^2$ pixel (chosen to roughly match the optical resolution). These values can be compared with the theoretical sensitivity in the photon shot noise limit, given by~\cite{Barry2020}
\begin{equation}\label{eq:sens}
\eta_B=\frac{4}{3\sqrt{3}\gamma_e}\frac{\Delta\nu}{{\cal C} \sqrt{\cal R}}  \end{equation}
where $\Delta\nu$ is the ODMR linewidth (FWHM), $\cal C$ the ODMR contrast, and $\cal R$ the average photon detection rate (taking the dead times into account). Using the parameters extracted from the experiments, we find a per-pixel sensitivity of $\eta_B\approx192$ and $34\,\mu$T/$\sqrt{\rm Hz}$ with the LED and LD, respectively, reasonably close to the actual sensitivity observed indicating the measurement is shot-noise limited (the discrepancy is attributed to the non-optimal sampling of the ODMR spectrum). We note that the measurements reported in Fig.~\ref{fig3} were taken with the system sitting on a regular desk with normal ambient lighting, as pictured in Fig.~\ref{fig1}(c).   

From these comparative tests we conclude that the LD significantly outperforms the LED in this instance. This is because of the small field of view required to image our test sample [$\sim(500\,\mu$m$)^2$] allowing the LD spot size to be optimised accordingly to maximise the power density and hence the sensitivity. Nevertheless, the LED is a suitable option when fields of view at or larger than its minimum spot size ($\sim2.5$\,mm here) are required. A dual illumination system could also be easily incorporated in the sensing head, e.g. LD for NV excitation and LED to facilitate sample positioning. 

The magnetic sensitivity could be improved in several ways, as informed by Eq.~\ref{eq:sens}. First, NV layers formed by optimal nitrogen doping during chemical vapour deposition growth typically yield a more favourable $\Delta\nu/\sqrt{\cal R}$ ratio than for the HPHT diamond used here~\cite{Kleinsasser2016,Healey2020}. Second, the thickness of the NV layer should be increased to boost $\cal R$ so long as the spatial resolution does not become degraded as a result~\cite{Scholten2021}. For instance, here we could have used a 2-$\mu$m-thick NV layer instead of the 500-nm layer. Lastly, ${\cal C}$ can be increased further (e.g. to about 3\% with the diamond used) using a higher excitation power density ($\gtrsim100$\,W/cm$^2$). This will require a higher power laser if the field of view is to be preserved. With the above improvements combined, a per-pixel sensitivity well below $1\,\mu$T/$\sqrt{\rm Hz}$ should be achievable (translating into a shorter acquisition time to reach a similar noise level), with modest impact on the portability of the system.   

The spatial resolution in the magnetic images results from a convolution between the optical resolution (3.9\,$\mu$m in the current conditions, as determined above) and the effect of the finite standoff $d$ between the sample and the NV sensing layer~\cite{Abrahams2021}. By fitting a pair of line profiles across perpendicular edges of the magnetic film [Fig.~\ref{fig3}(h)] using the model described in Refs.~\cite{Hingant2015,Abrahams2021}, we deduce a standoff of $d\approx6.6\pm1.2\,\mu$m. This indicates that the magnetic resolution is here dominated by the standoff effect rather than the optical resolution. Minimising the relative angle between sample and diamond using a goniometric stage was shown to improve standoff, with $d\approx1.6\,\mu$m achieved in Ref.~\cite{Abrahams2021}, and could be incorporated in our system. Simple procedures to clean the diamond surface after each use of the sensing head will also need to be developed. In addition, the optical resolution could be improved by reducing aberrations, e.g. using a thinner diamond~\cite{Nishimura2024}. With these improvements, magnetic images with a $2\,\mu$m resolution should be readily attainable in routine with a portable QDM system.

Finally, we note that the size, weight and power footprint of our current system could be reduced further, especially the control unit where a lower-power MW amplifier could be employed (e.g. Mini-Circuits ZRL-3500). The lower output MW power could be offset by a MW loop antenna with a smaller diameter, though at the cost of a reduced maximum field of view. 

In conclusion, we demonstrated a compact and portable version of a quantum diamond microscope, capable of quantitative magnetic field imaging outside a laboratory environment. The performance of our current system was characterised, and future improvements were outlined. Our device may find use in situations where taking the sample to be measured to a specialist lab is impractical or undesirable, for instance for rapid screening of rock samples in the field.
 
\section*{Supplementary Material}
See supplementary material for additional diamond sample details, additional photographs of the setup, and a full list of components.

\begin{acknowledgments}
The authors thank S.C. Scholten for assistance with ODMR data processing and A.J. Healey with diamond cleaning. I.O.R. is supported by an Australian Government Research Training Program Scholarship. 
This work was supported by the Australian Research Council (ARC) through grants FT200100073, DP220100178 and DE230100192. 
%S.C.S gratefully acknowledges the support of an Ernst and Grace Matthaei scholarship.
P.R. acknowledges support through an RMIT University Vice-Chancellor’s Research Fellowship. 
\end{acknowledgments}

\section*{Data Availability Statement}

The data that supports the findings of this study are available within the article and its supplementary material.

\section*{Author Declarations}

The authors have no conflicts to disclose.

\bibliography{refs}

\clearpage

\onecolumngrid

\begin{center}
\textbf{\large Supplementary Material}
\end{center}

\setcounter{equation}{0}
\setcounter{section}{0}
\setcounter{figure}{0}
\setcounter{table}{0}
\setcounter{page}{1}
\makeatletter
\renewcommand{\theequation}{S\arabic{equation}}
\renewcommand{\thefigure}{S\arabic{figure}}

\section{Diamond sensor details\label{sec:diamond}}

The diamond sensors used in this work were made from $4\,{\rm mm}\times 4\,{\rm mm}\times 400\,\mu{\rm m}$ type-Ib, single-crystal diamond substrates grown by high-pressure, high-temperature (HPHT) synthesis, with $\{100\}$-oriented polished faces, purchased from Chenguang Machinery \& Electric Equipment. To form a dense NV layer near the surface, the diamonds were implanted with $2$\,MeV Sb ions with a dose of $2\times 10^{11}$ ions/cm$^2$. Full cascade Stopping and Range of Ions in Matter (SRIM) Monte Carlo simulations indicate that the distribution of created vacancies extend to about 500-600 nm below the surface, with a peak at $\approx400$\,nm. Following implantation, the diamond was annealed at $900$\,$^\circ$C for 4 hours in a vacuum of $\sim10^{-5}$~Torr to form the NV centers. After annealing, the plate was acid cleaned ($30$\,minutes in a boiling mixture of sulphuric acid and sodium nitrate).

The same diamond was used throughout the main text except in Fig. 2(e) where a different diamond (though nominally identical) was employed to determine the optical resolution.

\section{Setup details}

This section provides further details about the device presented in the main text, with the aim of facilitating replication.

\begin{figure*}[b!]
    \centering
    \includegraphics[width=8cm]{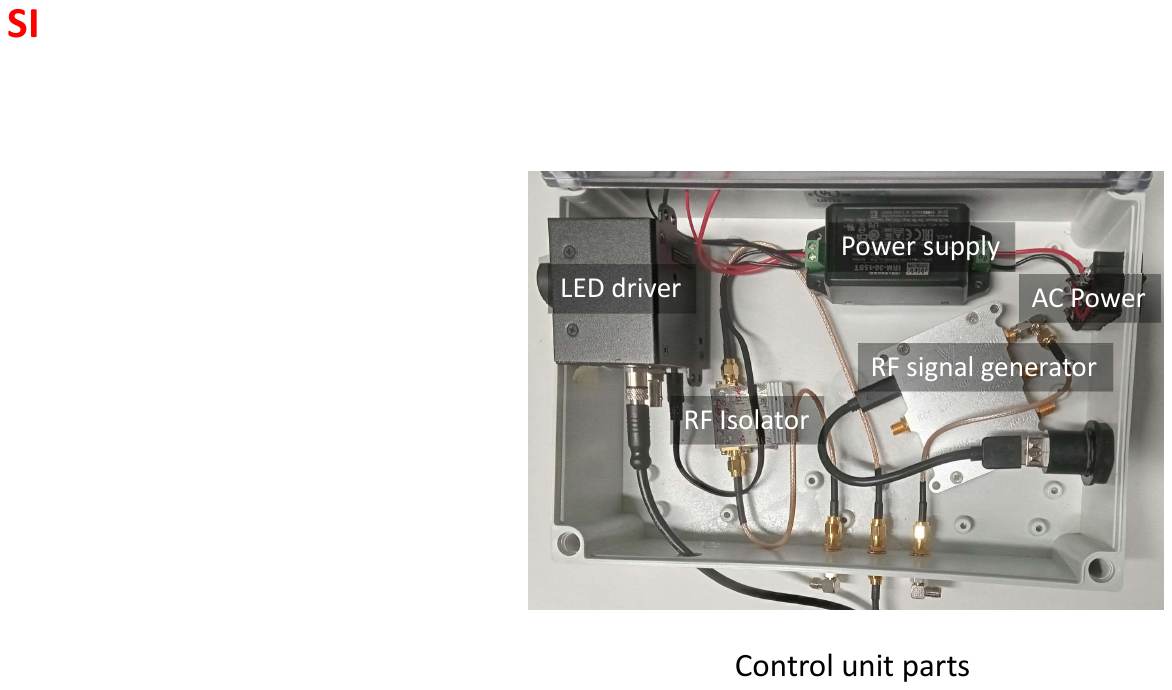}
    \caption{Photograph showing the components inside the control unit, for the LED version.}
    \label{SI_fig2}
\end{figure*}

The inside of the control unit is shown in Fig. \ref{SI_fig2}, and the components listed in Table \ref{tab:control unit}. The typical retail price of each component is also indicated. A MW isolator is included to protect the amplifier. A 30\,W, 15\,V AC/DC converter supplies power to the amplifier, fan, and LED controller. For the laser diode version (not shown), the LED Controller is replaced with the laser driver (listed in Table \ref{tab:light source}) and an appropriate AC/DC converter is added. Not shown in the photograph in Fig. \ref{SI_fig2} (but visible in Fig. 1(c) of the main text) is the microwave (MW) amplifier which is placed on the lid of the enclosing box, and the fan attached to the amplifier to prevent overheating.  

\begin{table*}
\begin{tabular}{|l|l|c|}
\hline
{\bf Description}  & {\bf Item reference} & {\bf Price (US\$)}  \\ 
\hline
Plastic box & Bud Industries, PIP-11775-C  & 55.69 \\
\hline
MW signal generator & Windfreak, SynthNV Pro & 1,199.00  \\ 
\hline
MW amplifier & Mini-Circuits, ZHL-42+ & 1,132.18 \\ 
\hline
MW isolator & UIY, UIYBCI3038A2T6SF & 360.00 \\ 
\hline
Fan for amplifier & Sanyo Denki, 109R0824G4021 & 51.17 \\ 
\hline
AC/DC converter & Mean Well, IRM-30-15ST & 19.49 \\ 
\hline
AC power connector & TE Connectivity, 1609112-3 & 18.31 \\ 
\hline
LED controller power cable & Tensility International Corp, CA-2193 & 4.14 \\ 
\hline
USB C to USB A connector & Adafruit Industries, 4259 & 9.47 \\ 
\hline
USB A to USB C cable & Adafruit Industries, 4472 & 4.69 \\ 
\hline
SMA to SMA connectors & Cinch Connectivity, 142-0901-401 (3x) & 55.14 \\ 
\hline
SMA to SMA coaxial cables & Cinch Connectivity, 415-0029-MM500 (5x) & 86.35 \\ 
\hline
  & Cinch Connectivity, 415-0029-036 & 19.01 \\ 
\hline
SMA 'L' connectors & TE Connectivity Linx, CONSMA010 (5x) & 26.50 \\ 
\hline
SMA to SMB coaxial cable & Taoglas Limited, CAB.0101 & 7.90 \\ 
\hline
\textbf{Total} &  & \textbf{3,049.04} \\
\hline
\end{tabular}
\caption{List of components contained in the control unit (excluding light source controller, which is listed in Table \ref{tab:light source}), and their typical retail price.
}	
\label{tab:control unit}
\end{table*}

\begin{figure*}[b!]
    \centering
    \includegraphics[width=15cm]{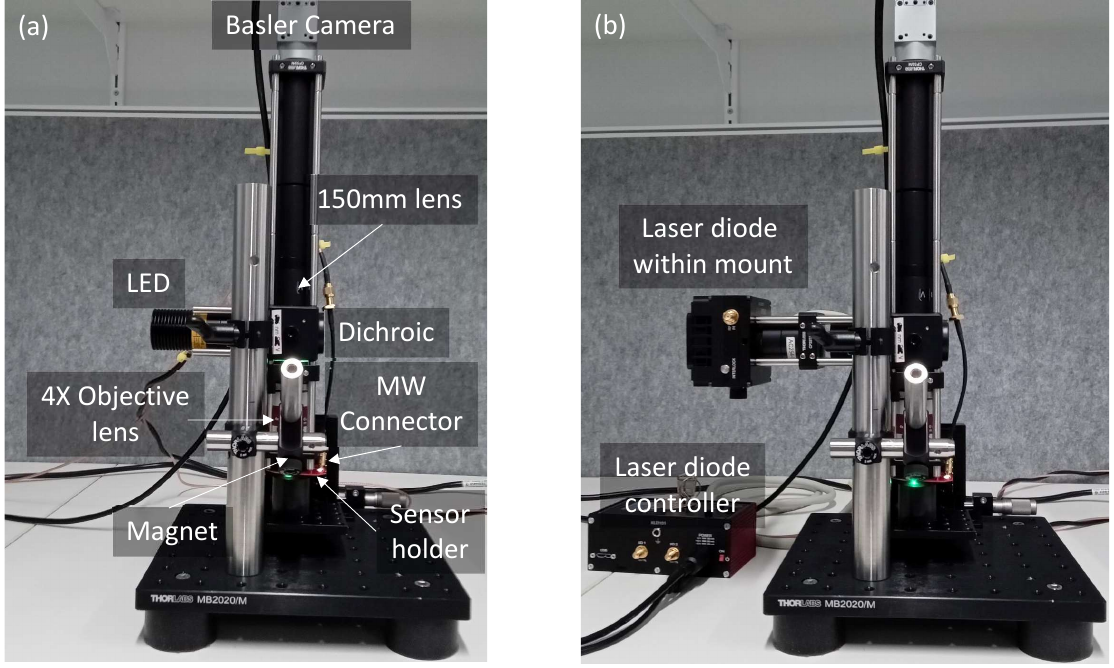}
    \caption{(a,b) Photograph showing the system with (a) LED and (b) laser diode, as the excitation source.}
    \label{SI_fig1}
\end{figure*}

Photographs of the microscopy setup (sensor head and sample stage on breadboard) are shown in Fig. \ref{SI_fig1}, taken from a different viewpoint to Fig. 1(c) of the main text to better show the light source. Fig. \ref{SI_fig1}(a) and \ref{SI_fig1}(b) show the version with the LED and laser diode, respectively. The components are listed in Table \ref{tab:sensor head}, except for the components specific to each light source which are listed separately in Table \ref{tab:light source}.

As mentioned in the main text, the only element that is not commercially available is the custom printed circuit board (PCB) holding the diamond. To connect the PCB to the optical system, a Thorlabs internal locking ring (SM1NT) was glued to the PCB and coupled to a lens tube through an external thread adapter (SM1T2). 

The total approximate cost of the entire system is indicated in Table \ref{tab:light source} for the two versions, not including the cost of the diamond sensor, custom PCB, laptop, and software.    

\begin{table*}[b!]
\begin{tabular}{|l|l|c|}
\hline
{\bf Description}  & {\bf Item reference} & {\bf Price (US\$)}  \\ 
\hline
Camera & Basler, acA2040-90um Mono USB 3.0 & 1,430.00 \\ 
\hline
Dichroic beam splitter (640 nm LP) & Semrock, FF640-FDi01-25x36 & 390.00 \\ 
\hline
Cage cube mount for dichroic & Thorlabs, CM1-DCH & 192.44  \\ 
\hline
Z-axis translation mount for objective & Thorlabs, SM1ZA & 220.20 \\ 
\hline
Objective lens & Thorlabs, TL4X-SAP & 2,308.01\\ 
\hline
Cage plates & Thorlabs, CP33T + CP36 + CP4S-SM1 & 72.23 \\ 
\hline
Filters (600 nm SP + 650 nm LP) & Thorlabs, FESH0600 + FELH0650 & 287.62 \\
\hline
Tube lens $f=150$\,mm & Thorlabs, AC254-150-B & 103.65 \\ 
\hline
Cage construction rods & Thorlabs, ER4 (4x) + ER3 (4x) + ER2 (8x) + ER1 (4x) & 132.32 \\ 
\hline
Lens tubes for optics & Thorlabs, SM1L30 + SM1L20 + SM1L10 (2x) + SM1L05 (3x) + SM1L03 & 131.89 \\ 
\hline
Adapters for PCB & Thorlabs, SM1NT + SM1T2 & 31.23 \\ 
\hline
SMB-PCB connector & JAE Electronics, SMBR004D00  & 3.00 \\ 
\hline
Neodymium magnet & AMF Magnetics, AMFYPC16 (threaded pot) + D-D18H15-N48 (disc)  & 16.03 \\ 
\hline
Construction posts &  Thorlabs, RS100P4 + RS150P4 + TR75 (2x) & 77.26  \\ 
\hline
Clamp to hold cage &  Thorlabs, CH1030 & 96.15  \\ 
\hline
Right angle clamps for magnet & Thorlabs, RA90 + RA90RS & 42.01 \\ 
\hline
Breadboard + rubber feet & Thorlabs, MB2020 + AV4 (4x) + AV3 (for sample) & 214.39  \\ 
\hline
XYZ translation stage for sample & Newport, M-460A-XYZ & 884.00 \\ 
\hline
Right angle bracket for sample stage & Thorlabs, AB90A & 30.58  \\
\hline
Plate for sample & Thorlabs, RB13P1 & 60.59 \\ 
\hline
\textbf{Total} &  & \textbf{6,723.60} \\ 
\hline
\end{tabular}
\caption{List of components making up the microscopy setup (excluding light source, which is listed in Table \ref{tab:light source}).
}	
\label{tab:sensor head}
\end{table*}

\begin{table*}[b!]
\begin{tabular}{|l|l|c|}
\hline
{\bf Description (LED version)}  & {\bf Item reference} & {\bf Price (US\$)}  \\ 
\hline
Mounted LED & Thorlabs, M530L4 & 325.49  \\ 
\hline
LED driver & Thorlabs, LEDD1B & 355.18  \\ 
\hline
Collimating lens $f=20$\,mm & Thorlabs, ACL2520U-A & 33.85  \\ 
\hline
{\bf Total} & & {\bf 714.52}\\ 
\hline
{\bf Final total (inc. all components)} & & {\bf 10,487.16}\\ 
\hline
\hline
{\bf Description (Laser diode version)}  & {\bf Item reference} & {\bf Price (US\$)}  \\ 
\hline
Laser diode & Thorlabs, L520P50 & 76.61 \\
\hline
Mount with integrated TEC & Thorlabs, LDM9T & 937.13 \\
\hline
Laser diode driver & Thorlabs, KLD101 & 973.85 \\
\hline
AC/DC power supply & Thorlabs, TPS002 & 128.29 \\
\hline
Collimating lens adapter & Thorlabs, S1TM12 & 26.73 \\
\hline
Collimating lens $f=8$\,mm & Thorlabs, C240TMD-A & 97.49  \\
\hline
Widefield lens $f=75$\,mm & Thorlabs, AC254-075-A & 89.10  \\
\hline
{\bf Total} & & {\bf 2,329.20}\\
\hline
{\bf Final total (inc. all components)} & & {\bf 12,101.84}\\ 
\hline
\end{tabular}
\caption{List of components specific to the LED or laser diode versions, with the total price of the entire system indicated.}	
\label{tab:light source}
\end{table*}

\end{document}